\documentstyle[sprocl]{article}
\def\Journal#1#2#3#4{{#1} {\bf #2}, #3 (#4)}

\def\NCA{\em Nuovo Cimento}
\def\NIM{\em Nucl. Instrum. Methods}
\def\NIMA{{\em Nucl. Instrum. Methods} A}
\def\NPB{{\em Nucl. Phys.} B}
\def\PLB{{\em Phys. Lett.}  B}
\def\PRL{\em Phys. Rev. Lett.}
\def\PRD{{\em Phys. Rev.} D}
\def\ZPC{{\em Z. Phys.} C}
\def\st{\scriptstyle}
\def\sst{\scriptscriptstyle}
\def\mco{\multicolumn}
\def\epp{\epsilon^{\prime}}
\def\vep{\varepsilon}
\def\ra{\rightarrow}
\def\ppg{\pi^+\pi^-\gamma}
\def\vp{{\bf p}}
\def\ko{K^0}
\def\kb{\bar{K^0}}
\def\al{\alpha}
\def\ab{\bar{\alpha}}
\def\be{\begin{equation}}
\def\ee{\end{equation}}
\def\bea{\begin{eqnarray}}
\def\eea{\end{eqnarray}}
\def\CPbar{\hbox{{\rm CP}\hskip-1.80em{/}}}

\begin{document}

\vspace*{-3.3cm}
\begin{center}
\begin{footnotesize}\baselineskip 10 pt\noindent
Proc.~~{\it Large Scale Structure: Tracks and Traces},~~Potsdam, Germany,
Sept\,15--19, 1997 \\
World Scientific Publ., Singapore, 1998, \\
eds.~V.~M\"uller, S.\,Gottl\"ober, J.P.\,M\"ucket, \& J.\,Wambsganss
\end{footnotesize}
\end{center}
\vspace*{1.6cm}

\title{CURRENT STATUS OF THE ACO CLUSTER REDSHIFT COMPILATION}

\author{ H. ANDERNACH }

\address{Depto.~de Astronom\'\i a, IFUG, 36000 Guanajuato, Gto., Mexico}

\author{ E. TAGO }

\address{Tartu Observatory, EE--2444 T\~oravere, Estonia}


\maketitle\abstracts{
We present an update of our compilation of measured redshifts
of galaxy clusters in the all-sky Abell catalog.
In the last 7 years the number of ACO clusters with measured $z$
has doubled to now $\sim$2100, but still $\sim$56\% of these are based
on only 1~or~2 measured member galaxies.
Our October 1997 version gives 2247 redshifts (including components or
line-of-sight superpositions) for 2114 distinct A- and S-clusters.
Velocity dispersions are listed for 536 different ACO clusters (613 
subclusters) and the median is 695 km/s.
We mention some applications of our compilation for determining the
large-scale structure of the nearby Universe.
}

\section{Introduction and Scope of this Work}

Since the publication of ACO's all-sky cluster catalog~\cite{aco89}
(with 822 redshifts) no redshift compilation of the full sample appeared in
print. We present an update of our catalog of measured redshifts for
both A- and S-clusters we maintain since 1989~\cite{and91,ats95}.
Different from most previous compilations, we systematically scan the
literature for any galaxy redshifts in the direction of ACO clusters.
Apart from compilations like ZCAT and SRC (cf.~\cite{ats95}), recent
redshift surveys (e.g.\ LCRS~\cite{lcrs} or APMBG~\cite{apmbg}) proved to be 
rich in ACO cluster redshifts, though these galaxy--cluster associations 
have not been reported in literature.
We include redshifts within a factor of 4 of the currently best
photometric estimate z$_{est}$~\cite{pw92}. We attach a special flag to
those with $|$log(z$_{obs}$/z$_{est})| >$ 0.3
and we include these entries only if the galaxies
are within $\sim$0.6 Abell radii (at the galaxy redshift).
Otherwise we include galaxies out to about 1 Abell radius.
The format of our compilation has been described previously~\cite{ats95}.
The bibliography has now 351 references and grows by about 25 papers
per year.

\section{Current Status and Exploratory Analysis}

Since three years ago\,\cite{ats95} the number of clusters with
measured redshifts has increased by $\sim$30\%, and the number of
known velocity dispersions by $\sim$60\%. The latter
is largely due to our efforts to collect literature data on individual
galaxies and calculate velocity dispersions by ourselves.
Our Oct.~1997 version gives redshifts for 2114 distinct
(1706 A- and 408 S-) clusters, or 2247 entries including subclusters or
line-of-sight superpositions.
Of these, $\sim$2150 entries (for $\sim$2050 different ACO names)
are within a factor of 2 from z$_{est}$.
In Fig.\,1 the number of clusters (N$_{cl}$) with redshifts
for at least N$_{z}$ member galaxies is plotted vs.~N$_{z}$,
for the ``SR91''~\cite{sr91} redshift compilation of northern ACO clusters,
as well as for our previous~\cite{ats95} and present full-sky compilations.
We give comments (e.g.~on additional or alternative redshifts or velocity
dispersions) for 865 or 38\% of all entries.

Since SR91~\cite{sr91} the fraction of clusters with ``reliable'' redshifts
(i.e. N$_{z}\ge$3) has risen from $\sim$32\% to $\sim$44\% now,
partly due to our efforts to merge data from different sources,
especially for low N$_{z}$. Such ``reliable'' redshifts now exist for 860
(670 A- and 190 S-) clusters.
There are now 100 clusters with N$_{z}\ge$50, the typical minimum
for studies of individual cluster dynamics. The total number of galaxies
involved (i.e.\ the sum of all N$_{z}$) is 22,200~~(11,700 for A-, 7020 
for ACO-, and 3480 for S-clusters),
neglecting some overlap for a few clusters.

Whenever individual redshifts are available for N$\ge$5 galaxies, we combine
various sources and calculate a velocity dispersion~$\sigma_V$.
We quote $\sigma_V$ for 536 different ACO clusters (400 A- and 136 S-clusters)
for a total of 613 subclusters.
A rather non-Gaussian (top-hat) main peak (FWHM $\sim$650 km/s) is followed by
a long tail out to 2000 km/s. Experience shows that some clusters in this
tail will turn out to be line-of-sight superpositions or cluster mergers.

The median $\sigma_V$ for all 613 (451 A- and 162 S-) clusters and subclusters
is 695 km/s (Fig.\,2), close to the median of 428 values based on clusters
with N$_{z}\geq$10 (705 km/s).
For the A-clusters alone the median~$\sigma_V$ is somewhat higher
(700 km/s) than for the S-clusters (657 km/s).
We suppose the reason for these values being smaller
than previously reported~\cite{ats95} is a more careful clipping
of outlyers and separation of subclusters.

There is a noticable increase of the velocity dispersion $\sigma_V$
with cluster redshift, probably due to both
a higher fraction of richer clusters and a higher
line-of-sight contamination with interloping galaxies at higher z.
There is also a trend for $\sigma_V$ to increase with
cluster richness R. We find a median $\sigma_V$ of 641, 842, and 1200
km/s for R$=$0$-$1, 2$-$3, and 4$-$5, respectively.
We do not see any dependence of $\sigma_V$ on Bautz-Morgan type.

In the course of our work we came across a new pair of identical 
clusters (A1664=A3541) and another five probably identical pairs or 
at least tightly connected clusters (A1681/A1683, A2122/A2124, 
A3389/S\,585, A3896/S1051, S\,622/S\,624).


\begin{minipage}[b]{.46\linewidth}


\setlength{\unitlength}{0.240900pt}
\ifx\plotpoint\undefined\newsavebox{\plotpoint}\fi
\sbox{\plotpoint}{\rule[-0.200pt]{0.400pt}{0.400pt}}%
\begin{picture}(900,720)(250,0)
\font\gnuplot=cmr10 at 10pt
\gnuplot
\sbox{\plotpoint}{\rule[-0.200pt]{0.400pt}{0.400pt}}%
\put(220.0,113.0){\rule[-0.200pt]{4.818pt}{0.400pt}}
\put(198,113){\makebox(0,0)[r]{1}}
\put(816.0,113.0){\rule[-0.200pt]{4.818pt}{0.400pt}}
\put(220.0,164.0){\rule[-0.200pt]{2.409pt}{0.400pt}}
\put(826.0,164.0){\rule[-0.200pt]{2.409pt}{0.400pt}}
\put(220.0,193.0){\rule[-0.200pt]{2.409pt}{0.400pt}}
\put(826.0,193.0){\rule[-0.200pt]{2.409pt}{0.400pt}}
\put(220.0,214.0){\rule[-0.200pt]{2.409pt}{0.400pt}}
\put(826.0,214.0){\rule[-0.200pt]{2.409pt}{0.400pt}}
\put(220.0,230.0){\rule[-0.200pt]{2.409pt}{0.400pt}}
\put(826.0,230.0){\rule[-0.200pt]{2.409pt}{0.400pt}}
\put(220.0,244.0){\rule[-0.200pt]{2.409pt}{0.400pt}}
\put(826.0,244.0){\rule[-0.200pt]{2.409pt}{0.400pt}}
\put(220.0,255.0){\rule[-0.200pt]{2.409pt}{0.400pt}}
\put(826.0,255.0){\rule[-0.200pt]{2.409pt}{0.400pt}}
\put(220.0,265.0){\rule[-0.200pt]{2.409pt}{0.400pt}}
\put(826.0,265.0){\rule[-0.200pt]{2.409pt}{0.400pt}}
\put(220.0,273.0){\rule[-0.200pt]{2.409pt}{0.400pt}}
\put(826.0,273.0){\rule[-0.200pt]{2.409pt}{0.400pt}}
\put(220.0,281.0){\rule[-0.200pt]{4.818pt}{0.400pt}}
\put(198,281){\makebox(0,0)[r]{10}}
\put(816.0,281.0){\rule[-0.200pt]{4.818pt}{0.400pt}}
\put(220.0,332.0){\rule[-0.200pt]{2.409pt}{0.400pt}}
\put(826.0,332.0){\rule[-0.200pt]{2.409pt}{0.400pt}}
\put(220.0,361.0){\rule[-0.200pt]{2.409pt}{0.400pt}}
\put(826.0,361.0){\rule[-0.200pt]{2.409pt}{0.400pt}}
\put(220.0,382.0){\rule[-0.200pt]{2.409pt}{0.400pt}}
\put(826.0,382.0){\rule[-0.200pt]{2.409pt}{0.400pt}}
\put(220.0,398.0){\rule[-0.200pt]{2.409pt}{0.400pt}}
\put(826.0,398.0){\rule[-0.200pt]{2.409pt}{0.400pt}}
\put(220.0,412.0){\rule[-0.200pt]{2.409pt}{0.400pt}}
\put(826.0,412.0){\rule[-0.200pt]{2.409pt}{0.400pt}}
\put(220.0,423.0){\rule[-0.200pt]{2.409pt}{0.400pt}}
\put(826.0,423.0){\rule[-0.200pt]{2.409pt}{0.400pt}}
\put(220.0,433.0){\rule[-0.200pt]{2.409pt}{0.400pt}}
\put(826.0,433.0){\rule[-0.200pt]{2.409pt}{0.400pt}}
\put(220.0,441.0){\rule[-0.200pt]{2.409pt}{0.400pt}}
\put(826.0,441.0){\rule[-0.200pt]{2.409pt}{0.400pt}}
\put(220.0,449.0){\rule[-0.200pt]{4.818pt}{0.400pt}}
\put(198,449){\makebox(0,0)[r]{100}}
\put(816.0,449.0){\rule[-0.200pt]{4.818pt}{0.400pt}}
\put(220.0,499.0){\rule[-0.200pt]{2.409pt}{0.400pt}}
\put(826.0,499.0){\rule[-0.200pt]{2.409pt}{0.400pt}}
\put(220.0,529.0){\rule[-0.200pt]{2.409pt}{0.400pt}}
\put(826.0,529.0){\rule[-0.200pt]{2.409pt}{0.400pt}}
\put(220.0,550.0){\rule[-0.200pt]{2.409pt}{0.400pt}}
\put(826.0,550.0){\rule[-0.200pt]{2.409pt}{0.400pt}}
\put(220.0,566.0){\rule[-0.200pt]{2.409pt}{0.400pt}}
\put(826.0,566.0){\rule[-0.200pt]{2.409pt}{0.400pt}}
\put(220.0,580.0){\rule[-0.200pt]{2.409pt}{0.400pt}}
\put(826.0,580.0){\rule[-0.200pt]{2.409pt}{0.400pt}}
\put(220.0,591.0){\rule[-0.200pt]{2.409pt}{0.400pt}}
\put(826.0,591.0){\rule[-0.200pt]{2.409pt}{0.400pt}}
\put(220.0,601.0){\rule[-0.200pt]{2.409pt}{0.400pt}}
\put(826.0,601.0){\rule[-0.200pt]{2.409pt}{0.400pt}}
\put(220.0,609.0){\rule[-0.200pt]{2.409pt}{0.400pt}}
\put(826.0,609.0){\rule[-0.200pt]{2.409pt}{0.400pt}}
\put(220.0,617.0){\rule[-0.200pt]{4.818pt}{0.400pt}}
\put(198,617){\makebox(0,0)[r]{1000}}
\put(816.0,617.0){\rule[-0.200pt]{4.818pt}{0.400pt}}
\put(220.0,667.0){\rule[-0.200pt]{2.409pt}{0.400pt}}
\put(826.0,667.0){\rule[-0.200pt]{2.409pt}{0.400pt}}
\put(220.0,697.0){\rule[-0.200pt]{2.409pt}{0.400pt}}
\put(826.0,697.0){\rule[-0.200pt]{2.409pt}{0.400pt}}
\put(220.0,113.0){\rule[-0.200pt]{0.400pt}{4.818pt}}
\put(220,68){\makebox(0,0){1}}
\put(220.0,677.0){\rule[-0.200pt]{0.400pt}{4.818pt}}
\put(289.0,113.0){\rule[-0.200pt]{0.400pt}{2.409pt}}
\put(289.0,687.0){\rule[-0.200pt]{0.400pt}{2.409pt}}
\put(329.0,113.0){\rule[-0.200pt]{0.400pt}{2.409pt}}
\put(329.0,687.0){\rule[-0.200pt]{0.400pt}{2.409pt}}
\put(357.0,113.0){\rule[-0.200pt]{0.400pt}{2.409pt}}
\put(357.0,687.0){\rule[-0.200pt]{0.400pt}{2.409pt}}
\put(380.0,113.0){\rule[-0.200pt]{0.400pt}{2.409pt}}
\put(380.0,687.0){\rule[-0.200pt]{0.400pt}{2.409pt}}
\put(398.0,113.0){\rule[-0.200pt]{0.400pt}{2.409pt}}
\put(398.0,687.0){\rule[-0.200pt]{0.400pt}{2.409pt}}
\put(413.0,113.0){\rule[-0.200pt]{0.400pt}{2.409pt}}
\put(413.0,687.0){\rule[-0.200pt]{0.400pt}{2.409pt}}
\put(426.0,113.0){\rule[-0.200pt]{0.400pt}{2.409pt}}
\put(426.0,687.0){\rule[-0.200pt]{0.400pt}{2.409pt}}
\put(438.0,113.0){\rule[-0.200pt]{0.400pt}{2.409pt}}
\put(438.0,687.0){\rule[-0.200pt]{0.400pt}{2.409pt}}
\put(448.0,113.0){\rule[-0.200pt]{0.400pt}{4.818pt}}
\put(448,68){\makebox(0,0){10}}
\put(448.0,677.0){\rule[-0.200pt]{0.400pt}{4.818pt}}
\put(517.0,113.0){\rule[-0.200pt]{0.400pt}{2.409pt}}
\put(517.0,687.0){\rule[-0.200pt]{0.400pt}{2.409pt}}
\put(557.0,113.0){\rule[-0.200pt]{0.400pt}{2.409pt}}
\put(557.0,687.0){\rule[-0.200pt]{0.400pt}{2.409pt}}
\put(586.0,113.0){\rule[-0.200pt]{0.400pt}{2.409pt}}
\put(586.0,687.0){\rule[-0.200pt]{0.400pt}{2.409pt}}
\put(608.0,113.0){\rule[-0.200pt]{0.400pt}{2.409pt}}
\put(608.0,687.0){\rule[-0.200pt]{0.400pt}{2.409pt}}
\put(626.0,113.0){\rule[-0.200pt]{0.400pt}{2.409pt}}
\put(626.0,687.0){\rule[-0.200pt]{0.400pt}{2.409pt}}
\put(641.0,113.0){\rule[-0.200pt]{0.400pt}{2.409pt}}
\put(641.0,687.0){\rule[-0.200pt]{0.400pt}{2.409pt}}
\put(654.0,113.0){\rule[-0.200pt]{0.400pt}{2.409pt}}
\put(654.0,687.0){\rule[-0.200pt]{0.400pt}{2.409pt}}
\put(666.0,113.0){\rule[-0.200pt]{0.400pt}{2.409pt}}
\put(666.0,687.0){\rule[-0.200pt]{0.400pt}{2.409pt}}
\put(676.0,113.0){\rule[-0.200pt]{0.400pt}{4.818pt}}
\put(676,68){\makebox(0,0){100}}
\put(676.0,677.0){\rule[-0.200pt]{0.400pt}{4.818pt}}
\put(745.0,113.0){\rule[-0.200pt]{0.400pt}{2.409pt}}
\put(745.0,687.0){\rule[-0.200pt]{0.400pt}{2.409pt}}
\put(785.0,113.0){\rule[-0.200pt]{0.400pt}{2.409pt}}
\put(785.0,687.0){\rule[-0.200pt]{0.400pt}{2.409pt}}
\put(814.0,113.0){\rule[-0.200pt]{0.400pt}{2.409pt}}
\put(814.0,687.0){\rule[-0.200pt]{0.400pt}{2.409pt}}
\put(836.0,113.0){\rule[-0.200pt]{0.400pt}{2.409pt}}
\put(836.0,687.0){\rule[-0.200pt]{0.400pt}{2.409pt}}
\put(220.0,113.0){\rule[-0.200pt]{148.394pt}{0.400pt}}
\put(836.0,113.0){\rule[-0.200pt]{0.400pt}{140.686pt}}
\put(220.0,697.0){\rule[-0.200pt]{148.394pt}{0.400pt}}
\put(145,360){\makebox(0,0){$N_{cl}$}}
\put(528,23){\makebox(0,0){$N_{z}$}}
\put(220.0,113.0){\rule[-0.200pt]{0.400pt}{140.686pt}}
\sbox{\plotpoint}{\rule[-0.600pt]{1.200pt}{1.200pt}}%
\put(706,632){\makebox(0,0)[r]{AT97}}
\put(728.0,632.0){\rule[-0.600pt]{15.899pt}{1.200pt}}
\put(220,676){\usebox{\plotpoint}}
\multiput(220.00,673.26)(1.111,-0.500){52}{\rule{2.971pt}{0.121pt}}
\multiput(220.00,673.51)(62.834,-31.000){2}{\rule{1.485pt}{1.200pt}}
\multiput(289.00,642.26)(0.678,-0.500){48}{\rule{1.955pt}{0.121pt}}
\multiput(289.00,642.51)(35.942,-29.000){2}{\rule{0.978pt}{1.200pt}}
\multiput(329.00,613.26)(0.719,-0.501){28}{\rule{2.068pt}{0.121pt}}
\multiput(329.00,613.51)(23.707,-19.000){2}{\rule{1.034pt}{1.200pt}}
\multiput(357.00,594.26)(1.144,-0.502){10}{\rule{3.060pt}{0.121pt}}
\multiput(357.00,594.51)(16.649,-10.000){2}{\rule{1.530pt}{1.200pt}}
\multiput(380.00,584.26)(1.109,-0.503){6}{\rule{3.000pt}{0.121pt}}
\multiput(380.00,584.51)(11.773,-8.000){2}{\rule{1.500pt}{1.200pt}}
\put(398,574.01){\rule{3.614pt}{1.200pt}}
\multiput(398.00,576.51)(7.500,-5.000){2}{\rule{1.807pt}{1.200pt}}
\multiput(413.00,571.25)(0.962,-0.509){2}{\rule{2.900pt}{0.123pt}}
\multiput(413.00,571.51)(6.981,-6.000){2}{\rule{1.450pt}{1.200pt}}
\multiput(426.00,565.25)(0.792,-0.509){2}{\rule{2.700pt}{0.123pt}}
\multiput(426.00,565.51)(6.396,-6.000){2}{\rule{1.350pt}{1.200pt}}
\put(438,557.01){\rule{2.409pt}{1.200pt}}
\multiput(438.00,559.51)(5.000,-5.000){2}{\rule{1.204pt}{1.200pt}}
\put(448,552.01){\rule{2.409pt}{1.200pt}}
\multiput(448.00,554.51)(5.000,-5.000){2}{\rule{1.204pt}{1.200pt}}
\put(458,547.51){\rule{1.927pt}{1.200pt}}
\multiput(458.00,549.51)(4.000,-4.000){2}{\rule{0.964pt}{1.200pt}}
\put(466,543.51){\rule{1.927pt}{1.200pt}}
\multiput(466.00,545.51)(4.000,-4.000){2}{\rule{0.964pt}{1.200pt}}
\put(474,539.01){\rule{1.927pt}{1.200pt}}
\multiput(474.00,541.51)(4.000,-5.000){2}{\rule{0.964pt}{1.200pt}}
\put(482,534.01){\rule{1.445pt}{1.200pt}}
\multiput(482.00,536.51)(3.000,-5.000){2}{\rule{0.723pt}{1.200pt}}
\put(488,529.51){\rule{1.686pt}{1.200pt}}
\multiput(488.00,531.51)(3.500,-4.000){2}{\rule{0.843pt}{1.200pt}}
\put(495,526.01){\rule{1.445pt}{1.200pt}}
\multiput(495.00,527.51)(3.000,-3.000){2}{\rule{0.723pt}{1.200pt}}
\put(501,522.51){\rule{1.204pt}{1.200pt}}
\multiput(501.00,524.51)(2.500,-4.000){2}{\rule{0.602pt}{1.200pt}}
\put(506,518.51){\rule{1.445pt}{1.200pt}}
\multiput(506.00,520.51)(3.000,-4.000){2}{\rule{0.723pt}{1.200pt}}
\put(512,515.51){\rule{1.204pt}{1.200pt}}
\multiput(512.00,516.51)(2.500,-2.000){2}{\rule{0.602pt}{1.200pt}}
\put(517,513.51){\rule{1.204pt}{1.200pt}}
\multiput(517.00,514.51)(2.500,-2.000){2}{\rule{0.602pt}{1.200pt}}
\put(522,511.01){\rule{0.964pt}{1.200pt}}
\multiput(522.00,512.51)(2.000,-3.000){2}{\rule{0.482pt}{1.200pt}}
\put(526,508.01){\rule{1.204pt}{1.200pt}}
\multiput(526.00,509.51)(2.500,-3.000){2}{\rule{0.602pt}{1.200pt}}
\put(531,506.01){\rule{0.964pt}{1.200pt}}
\multiput(531.00,506.51)(2.000,-1.000){2}{\rule{0.482pt}{1.200pt}}
\put(535,503.51){\rule{0.964pt}{1.200pt}}
\multiput(535.00,505.51)(2.000,-4.000){2}{\rule{0.482pt}{1.200pt}}
\put(539,500.51){\rule{0.964pt}{1.200pt}}
\multiput(539.00,501.51)(2.000,-2.000){2}{\rule{0.482pt}{1.200pt}}
\put(543,499.01){\rule{0.964pt}{1.200pt}}
\multiput(543.00,499.51)(2.000,-1.000){2}{\rule{0.482pt}{1.200pt}}
\put(547,497.51){\rule{0.723pt}{1.200pt}}
\multiput(547.00,498.51)(1.500,-2.000){2}{\rule{0.361pt}{1.200pt}}
\put(550,496.01){\rule{0.964pt}{1.200pt}}
\multiput(550.00,496.51)(2.000,-1.000){2}{\rule{0.482pt}{1.200pt}}
\put(554,494.01){\rule{0.723pt}{1.200pt}}
\multiput(554.00,495.51)(1.500,-3.000){2}{\rule{0.361pt}{1.200pt}}
\put(556.01,491){\rule{1.200pt}{0.964pt}}
\multiput(554.51,493.00)(3.000,-2.000){2}{\rule{1.200pt}{0.482pt}}
\put(560,487.01){\rule{0.964pt}{1.200pt}}
\multiput(560.00,488.51)(2.000,-3.000){2}{\rule{0.482pt}{1.200pt}}
\put(564,484.01){\rule{0.723pt}{1.200pt}}
\multiput(564.00,485.51)(1.500,-3.000){2}{\rule{0.361pt}{1.200pt}}
\put(566.01,481){\rule{1.200pt}{0.964pt}}
\multiput(564.51,483.00)(3.000,-2.000){2}{\rule{1.200pt}{0.482pt}}
\put(570,477.51){\rule{0.482pt}{1.200pt}}
\multiput(570.00,478.51)(1.000,-2.000){2}{\rule{0.241pt}{1.200pt}}
\put(572,475.01){\rule{0.723pt}{1.200pt}}
\multiput(572.00,476.51)(1.500,-3.000){2}{\rule{0.361pt}{1.200pt}}
\put(574.01,472){\rule{1.200pt}{0.964pt}}
\multiput(572.51,474.00)(3.000,-2.000){2}{\rule{1.200pt}{0.482pt}}
\put(578,469.01){\rule{0.723pt}{1.200pt}}
\multiput(578.00,469.51)(1.500,-1.000){2}{\rule{0.361pt}{1.200pt}}
\put(581,468.01){\rule{0.482pt}{1.200pt}}
\multiput(581.00,468.51)(1.000,-1.000){2}{\rule{0.241pt}{1.200pt}}
\put(582.01,466){\rule{1.200pt}{0.964pt}}
\multiput(580.51,468.00)(3.000,-2.000){2}{\rule{1.200pt}{0.482pt}}
\put(584.51,462){\rule{1.200pt}{0.964pt}}
\multiput(583.51,464.00)(2.000,-2.000){2}{\rule{1.200pt}{0.482pt}}
\put(588,458.51){\rule{0.482pt}{1.200pt}}
\multiput(588.00,459.51)(1.000,-2.000){2}{\rule{0.241pt}{1.200pt}}
\put(590,456.51){\rule{0.723pt}{1.200pt}}
\multiput(590.00,457.51)(1.500,-2.000){2}{\rule{0.361pt}{1.200pt}}
\put(595,454.01){\rule{0.964pt}{1.200pt}}
\multiput(595.00,455.51)(2.000,-3.000){2}{\rule{0.482pt}{1.200pt}}
\put(599,451.01){\rule{0.723pt}{1.200pt}}
\multiput(599.00,452.51)(1.500,-3.000){2}{\rule{0.361pt}{1.200pt}}
\put(593.0,458.0){\usebox{\plotpoint}}
\put(604,448.51){\rule{0.964pt}{1.200pt}}
\multiput(604.00,449.51)(2.000,-2.000){2}{\rule{0.482pt}{1.200pt}}
\put(608,447.01){\rule{0.482pt}{1.200pt}}
\multiput(608.00,447.51)(1.000,-1.000){2}{\rule{0.241pt}{1.200pt}}
\put(608.51,446){\rule{1.200pt}{0.723pt}}
\multiput(607.51,447.50)(2.000,-1.500){2}{\rule{1.200pt}{0.361pt}}
\put(612,442.51){\rule{0.482pt}{1.200pt}}
\multiput(612.00,443.51)(1.000,-2.000){2}{\rule{0.241pt}{1.200pt}}
\put(602.0,452.0){\usebox{\plotpoint}}
\put(617,440.51){\rule{0.482pt}{1.200pt}}
\multiput(617.00,441.51)(1.000,-2.000){2}{\rule{0.241pt}{1.200pt}}
\put(617.51,438){\rule{1.200pt}{0.964pt}}
\multiput(616.51,440.00)(2.000,-2.000){2}{\rule{1.200pt}{0.482pt}}
\put(621,435.01){\rule{0.241pt}{1.200pt}}
\multiput(621.00,435.51)(0.500,-1.000){2}{\rule{0.120pt}{1.200pt}}
\put(622,433.51){\rule{0.482pt}{1.200pt}}
\multiput(622.00,434.51)(1.000,-2.000){2}{\rule{0.241pt}{1.200pt}}
\put(624,432.01){\rule{0.723pt}{1.200pt}}
\multiput(624.00,432.51)(1.500,-1.000){2}{\rule{0.361pt}{1.200pt}}
\put(627,431.01){\rule{0.482pt}{1.200pt}}
\multiput(627.00,431.51)(1.000,-1.000){2}{\rule{0.241pt}{1.200pt}}
\put(629,430.01){\rule{0.482pt}{1.200pt}}
\multiput(629.00,430.51)(1.000,-1.000){2}{\rule{0.241pt}{1.200pt}}
\put(631,428.01){\rule{0.723pt}{1.200pt}}
\multiput(631.00,429.51)(1.500,-3.000){2}{\rule{0.361pt}{1.200pt}}
\put(632.01,427){\rule{1.200pt}{0.482pt}}
\multiput(631.51,428.00)(1.000,-1.000){2}{\rule{1.200pt}{0.241pt}}
\put(635,423.51){\rule{0.482pt}{1.200pt}}
\multiput(635.00,424.51)(1.000,-2.000){2}{\rule{0.241pt}{1.200pt}}
\put(635.01,422){\rule{1.200pt}{0.723pt}}
\multiput(634.51,423.50)(1.000,-1.500){2}{\rule{1.200pt}{0.361pt}}
\put(638,418.51){\rule{0.482pt}{1.200pt}}
\multiput(638.00,419.51)(1.000,-2.000){2}{\rule{0.241pt}{1.200pt}}
\put(638.01,416){\rule{1.200pt}{0.964pt}}
\multiput(637.51,418.00)(1.000,-2.000){2}{\rule{1.200pt}{0.482pt}}
\put(641,412.51){\rule{0.482pt}{1.200pt}}
\multiput(641.00,413.51)(1.000,-2.000){2}{\rule{0.241pt}{1.200pt}}
\put(641.01,410){\rule{1.200pt}{0.964pt}}
\multiput(640.51,412.00)(1.000,-2.000){2}{\rule{1.200pt}{0.482pt}}
\put(644,406.51){\rule{1.204pt}{1.200pt}}
\multiput(644.00,407.51)(2.500,-2.000){2}{\rule{0.602pt}{1.200pt}}
\put(647.51,405){\rule{1.200pt}{0.723pt}}
\multiput(646.51,406.50)(2.000,-1.500){2}{\rule{1.200pt}{0.361pt}}
\put(649.01,403){\rule{1.200pt}{0.482pt}}
\multiput(648.51,404.00)(1.000,-1.000){2}{\rule{1.200pt}{0.241pt}}
\put(650.01,401){\rule{1.200pt}{0.482pt}}
\multiput(649.51,402.00)(1.000,-1.000){2}{\rule{1.200pt}{0.241pt}}
\put(651.01,398){\rule{1.200pt}{0.723pt}}
\multiput(650.51,399.50)(1.000,-1.500){2}{\rule{1.200pt}{0.361pt}}
\put(654,395.01){\rule{0.482pt}{1.200pt}}
\multiput(654.00,395.51)(1.000,-1.000){2}{\rule{0.241pt}{1.200pt}}
\put(654.01,395){\rule{1.200pt}{0.482pt}}
\multiput(653.51,396.00)(1.000,-1.000){2}{\rule{1.200pt}{0.241pt}}
\put(655.51,392){\rule{1.200pt}{0.723pt}}
\multiput(654.51,393.50)(2.000,-1.500){2}{\rule{1.200pt}{0.361pt}}
\put(659,388.01){\rule{0.964pt}{1.200pt}}
\multiput(659.00,389.51)(2.000,-3.000){2}{\rule{0.482pt}{1.200pt}}
\put(661.01,387){\rule{1.200pt}{0.482pt}}
\multiput(660.51,388.00)(1.000,-1.000){2}{\rule{1.200pt}{0.241pt}}
\put(664,384.01){\rule{0.241pt}{1.200pt}}
\multiput(664.00,384.51)(0.500,-1.000){2}{\rule{0.120pt}{1.200pt}}
\put(663.01,384){\rule{1.200pt}{0.482pt}}
\multiput(662.51,385.00)(1.000,-1.000){2}{\rule{1.200pt}{0.241pt}}
\put(666,380.51){\rule{0.482pt}{1.200pt}}
\multiput(666.00,381.51)(1.000,-2.000){2}{\rule{0.241pt}{1.200pt}}
\put(666.01,380){\rule{1.200pt}{0.482pt}}
\multiput(665.51,381.00)(1.000,-1.000){2}{\rule{1.200pt}{0.241pt}}
\put(667.01,376){\rule{1.200pt}{0.964pt}}
\multiput(666.51,378.00)(1.000,-2.000){2}{\rule{1.200pt}{0.482pt}}
\put(670,372.51){\rule{0.482pt}{1.200pt}}
\multiput(670.00,373.51)(1.000,-2.000){2}{\rule{0.241pt}{1.200pt}}
\put(670.01,372){\rule{1.200pt}{0.482pt}}
\multiput(669.51,373.00)(1.000,-1.000){2}{\rule{1.200pt}{0.241pt}}
\put(673,368.51){\rule{0.482pt}{1.200pt}}
\multiput(673.00,369.51)(1.000,-2.000){2}{\rule{0.241pt}{1.200pt}}
\put(675,366.51){\rule{1.204pt}{1.200pt}}
\multiput(675.00,367.51)(2.500,-2.000){2}{\rule{0.602pt}{1.200pt}}
\put(678.01,366){\rule{1.200pt}{0.482pt}}
\multiput(677.51,367.00)(1.000,-1.000){2}{\rule{1.200pt}{0.241pt}}
\put(681,361.01){\rule{1.445pt}{1.200pt}}
\multiput(681.00,363.51)(3.000,-5.000){2}{\rule{0.723pt}{1.200pt}}
\put(685.01,359){\rule{1.200pt}{0.482pt}}
\multiput(684.51,360.00)(1.000,-1.000){2}{\rule{1.200pt}{0.241pt}}
\put(686.01,356){\rule{1.200pt}{0.723pt}}
\multiput(685.51,357.50)(1.000,-1.500){2}{\rule{1.200pt}{0.361pt}}
\put(614.0,444.0){\usebox{\plotpoint}}
\put(687.01,351){\rule{1.200pt}{0.482pt}}
\multiput(686.51,352.00)(1.000,-1.000){2}{\rule{1.200pt}{0.241pt}}
\put(690,347.01){\rule{1.204pt}{1.200pt}}
\multiput(690.00,348.51)(2.500,-3.000){2}{\rule{0.602pt}{1.200pt}}
\put(693.01,345){\rule{1.200pt}{0.723pt}}
\multiput(692.51,346.50)(1.000,-1.500){2}{\rule{1.200pt}{0.361pt}}
\put(696,341.01){\rule{0.723pt}{1.200pt}}
\multiput(696.00,342.51)(1.500,-3.000){2}{\rule{0.361pt}{1.200pt}}
\put(698.01,338){\rule{1.200pt}{0.964pt}}
\multiput(696.51,340.00)(3.000,-2.000){2}{\rule{1.200pt}{0.482pt}}
\put(701.51,332){\rule{1.200pt}{1.445pt}}
\multiput(699.51,335.00)(4.000,-3.000){2}{\rule{1.200pt}{0.723pt}}
\put(704.51,328){\rule{1.200pt}{0.964pt}}
\multiput(703.51,330.00)(2.000,-2.000){2}{\rule{1.200pt}{0.482pt}}
\put(708,323.51){\rule{0.964pt}{1.200pt}}
\multiput(708.00,325.51)(2.000,-4.000){2}{\rule{0.482pt}{1.200pt}}
\put(711.01,320){\rule{1.200pt}{0.964pt}}
\multiput(709.51,322.00)(3.000,-2.000){2}{\rule{1.200pt}{0.482pt}}
\put(714.51,315){\rule{1.200pt}{1.204pt}}
\multiput(712.51,317.50)(4.000,-2.500){2}{\rule{1.200pt}{0.602pt}}
\put(718.01,311){\rule{1.200pt}{0.964pt}}
\multiput(716.51,313.00)(3.000,-2.000){2}{\rule{1.200pt}{0.482pt}}
\put(720.51,305){\rule{1.200pt}{1.445pt}}
\multiput(719.51,308.00)(2.000,-3.000){2}{\rule{1.200pt}{0.723pt}}
\put(689.0,353.0){\usebox{\plotpoint}}
\put(724,294.51){\rule{1.686pt}{1.200pt}}
\multiput(724.00,297.51)(3.500,-6.000){2}{\rule{0.843pt}{1.200pt}}
\put(730.51,288){\rule{1.200pt}{1.445pt}}
\multiput(728.51,291.00)(4.000,-3.000){2}{\rule{1.200pt}{0.723pt}}
\put(735.01,281){\rule{1.200pt}{1.686pt}}
\multiput(732.51,284.50)(5.000,-3.500){2}{\rule{1.200pt}{0.843pt}}
\put(740.01,273){\rule{1.200pt}{1.927pt}}
\multiput(737.51,277.00)(5.000,-4.000){2}{\rule{1.200pt}{0.964pt}}
\multiput(747.24,265.11)(0.509,-0.113){2}{\rule{0.123pt}{1.900pt}}
\multiput(742.51,269.06)(6.000,-4.056){2}{\rule{1.200pt}{0.950pt}}
\put(750.51,255){\rule{1.200pt}{2.409pt}}
\multiput(748.51,260.00)(4.000,-5.000){2}{\rule{1.200pt}{1.204pt}}
\put(724.0,300.0){\rule[-0.600pt]{1.200pt}{1.204pt}}
\multiput(755.00,241.26)(0.685,-0.501){18}{\rule{2.014pt}{0.121pt}}
\multiput(755.00,241.51)(15.819,-14.000){2}{\rule{1.007pt}{1.200pt}}
\multiput(775.00,227.26)(0.465,-0.501){22}{\rule{1.500pt}{0.121pt}}
\multiput(775.00,227.51)(12.887,-16.000){2}{\rule{0.750pt}{1.200pt}}
\put(789.51,193){\rule{1.200pt}{5.059pt}}
\multiput(788.51,203.50)(2.000,-10.500){2}{\rule{1.200pt}{2.529pt}}
\put(791.01,164){\rule{1.200pt}{6.986pt}}
\multiput(790.51,178.50)(1.000,-14.500){2}{\rule{1.200pt}{3.493pt}}
\multiput(796.24,156.24)(0.500,-0.645){68}{\rule{0.121pt}{1.869pt}}
\multiput(791.51,160.12)(39.000,-47.120){2}{\rule{1.200pt}{0.935pt}}
\put(755.0,244.0){\rule[-0.600pt]{1.200pt}{2.650pt}}
\sbox{\plotpoint}{\rule[-0.500pt]{1.000pt}{1.000pt}}%
\put(706,587){\makebox(0,0)[r]{ATS95}}
\multiput(728,587)(20.756,0.000){4}{\usebox{\plotpoint}}
\put(794,587){\usebox{\plotpoint}}
\put(772,113){\usebox{\plotpoint}}
\multiput(772,113)(0.000,20.756){3}{\usebox{\plotpoint}}
\multiput(772,164)(-10.496,17.906){2}{\usebox{\plotpoint}}
\put(740.58,205.62){\usebox{\plotpoint}}
\put(727.06,221.00){\usebox{\plotpoint}}
\put(718.49,239.82){\usebox{\plotpoint}}
\multiput(717,244)(-13.143,16.064){0}{\usebox{\plotpoint}}
\put(707.40,257.01){\usebox{\plotpoint}}
\multiput(705,265)(-12.453,16.604){0}{\usebox{\plotpoint}}
\put(698.41,275.34){\usebox{\plotpoint}}
\multiput(697,281)(-1.592,20.694){0}{\usebox{\plotpoint}}
\put(694.79,295.45){\usebox{\plotpoint}}
\multiput(691,300)(-4.070,20.352){0}{\usebox{\plotpoint}}
\multiput(690,305)(-3.412,20.473){0}{\usebox{\plotpoint}}
\put(686.42,313.58){\usebox{\plotpoint}}
\multiput(685,315)(-16.889,12.064){0}{\usebox{\plotpoint}}
\multiput(678,320)(-5.034,20.136){0}{\usebox{\plotpoint}}
\put(675.54,329.84){\usebox{\plotpoint}}
\multiput(675,332)(-11.513,17.270){0}{\usebox{\plotpoint}}
\multiput(673,335)(-6.563,19.690){0}{\usebox{\plotpoint}}
\multiput(672,338)(-9.282,18.564){0}{\usebox{\plotpoint}}
\multiput(670,342)(-11.513,17.270){0}{\usebox{\plotpoint}}
\put(665.95,348.07){\usebox{\plotpoint}}
\multiput(664,351)(-9.282,18.564){0}{\usebox{\plotpoint}}
\multiput(663,353)(-6.563,19.690){0}{\usebox{\plotpoint}}
\multiput(662,356)(-17.798,10.679){0}{\usebox{\plotpoint}}
\multiput(657,359)(-9.282,18.564){0}{\usebox{\plotpoint}}
\multiput(656,361)(-14.676,14.676){0}{\usebox{\plotpoint}}
\put(653.82,363.92){\usebox{\plotpoint}}
\multiput(653,368)(-14.676,14.676){0}{\usebox{\plotpoint}}
\multiput(651,370)(-18.564,9.282){0}{\usebox{\plotpoint}}
\multiput(647,372)(-14.676,14.676){0}{\usebox{\plotpoint}}
\multiput(645,374)(-9.282,18.564){0}{\usebox{\plotpoint}}
\put(643.30,380.17){\usebox{\plotpoint}}
\multiput(643,382)(-9.282,18.564){0}{\usebox{\plotpoint}}
\multiput(641,386)(-3.412,20.473){0}{\usebox{\plotpoint}}
\multiput(640,392)(-14.676,14.676){0}{\usebox{\plotpoint}}
\multiput(638,394)(-16.604,12.453){0}{\usebox{\plotpoint}}
\put(633.51,397.16){\usebox{\plotpoint}}
\multiput(631,398)(-18.564,9.282){0}{\usebox{\plotpoint}}
\multiput(627,400)(-14.676,14.676){0}{\usebox{\plotpoint}}
\multiput(624,403)(-19.690,6.563){0}{\usebox{\plotpoint}}
\multiput(621,404)(-9.282,18.564){0}{\usebox{\plotpoint}}
\put(617.75,409.25){\usebox{\plotpoint}}
\multiput(617,410)(-17.270,11.513){0}{\usebox{\plotpoint}}
\multiput(614,412)(-14.676,14.676){0}{\usebox{\plotpoint}}
\multiput(612,414)(-7.708,19.271){0}{\usebox{\plotpoint}}
\multiput(610,419)(-18.564,9.282){0}{\usebox{\plotpoint}}
\multiput(608,420)(-16.604,12.453){0}{\usebox{\plotpoint}}
\put(603.43,423.28){\usebox{\plotpoint}}
\multiput(602,424)(-17.270,11.513){0}{\usebox{\plotpoint}}
\multiput(599,426)(-18.564,9.282){0}{\usebox{\plotpoint}}
\multiput(597,427)(-11.513,17.270){0}{\usebox{\plotpoint}}
\multiput(595,430)(-11.513,17.270){0}{\usebox{\plotpoint}}
\multiput(593,433)(-19.690,6.563){0}{\usebox{\plotpoint}}
\multiput(590,434)(-18.564,9.282){0}{\usebox{\plotpoint}}
\put(587.74,435.65){\usebox{\plotpoint}}
\multiput(586,440)(-19.690,6.563){0}{\usebox{\plotpoint}}
\multiput(583,441)(-17.798,10.679){0}{\usebox{\plotpoint}}
\multiput(578,444)(-17.270,11.513){0}{\usebox{\plotpoint}}
\multiput(575,446)(-19.690,6.563){0}{\usebox{\plotpoint}}
\put(571.83,447.26){\usebox{\plotpoint}}
\multiput(570,450)(-12.453,16.604){0}{\usebox{\plotpoint}}
\multiput(567,454)(-19.690,6.563){0}{\usebox{\plotpoint}}
\multiput(564,455)(-14.676,14.676){0}{\usebox{\plotpoint}}
\multiput(560,459)(-17.270,11.513){0}{\usebox{\plotpoint}}
\put(556.97,461.03){\usebox{\plotpoint}}
\multiput(554,464)(-18.564,9.282){0}{\usebox{\plotpoint}}
\multiput(550,466)(-17.270,11.513){0}{\usebox{\plotpoint}}
\multiput(547,468)(-20.136,5.034){0}{\usebox{\plotpoint}}
\multiput(543,469)(-20.136,5.034){0}{\usebox{\plotpoint}}
\put(538.82,470.14){\usebox{\plotpoint}}
\multiput(535,473)(-16.604,12.453){0}{\usebox{\plotpoint}}
\multiput(531,476)(-19.271,7.708){0}{\usebox{\plotpoint}}
\multiput(526,478)(-18.564,9.282){0}{\usebox{\plotpoint}}
\put(520.95,480.42){\usebox{\plotpoint}}
\multiput(517,482)(-20.352,4.070){0}{\usebox{\plotpoint}}
\multiput(512,483)(-17.270,11.513){0}{\usebox{\plotpoint}}
\put(502.41,489.15){\usebox{\plotpoint}}
\multiput(501,490)(-18.564,9.282){0}{\usebox{\plotpoint}}
\multiput(495,493)(-16.889,12.064){0}{\usebox{\plotpoint}}
\put(484.40,499.20){\usebox{\plotpoint}}
\multiput(482,500)(-17.601,11.000){0}{\usebox{\plotpoint}}
\multiput(474,505)(-19.434,7.288){0}{\usebox{\plotpoint}}
\put(465.77,508.11){\usebox{\plotpoint}}
\multiput(458,512)(-19.271,7.708){0}{\usebox{\plotpoint}}
\put(446.84,516.58){\usebox{\plotpoint}}
\put(428.28,525.86){\usebox{\plotpoint}}
\multiput(426,527)(-18.275,9.840){0}{\usebox{\plotpoint}}
\put(409.80,535.28){\usebox{\plotpoint}}
\put(390.97,543.91){\usebox{\plotpoint}}
\put(372.99,554.27){\usebox{\plotpoint}}
\multiput(357,564)(-15.759,13.508){2}{\usebox{\plotpoint}}
\multiput(329,588)(-15.814,13.442){3}{\usebox{\plotpoint}}
\multiput(289,622)(-18.618,9.174){3}{\usebox{\plotpoint}}
\put(220,656){\usebox{\plotpoint}}
\sbox{\plotpoint}{\rule[-0.200pt]{0.400pt}{0.400pt}}%
\put(706,542){\makebox(0,0)[r]{SR91}}
\put(728.0,542.0){\rule[-0.200pt]{15.899pt}{0.400pt}}
\put(220,559){\usebox{\plotpoint}}
\multiput(220.00,557.92)(0.785,-0.498){85}{\rule{0.727pt}{0.120pt}}
\multiput(220.00,558.17)(67.491,-44.000){2}{\rule{0.364pt}{0.400pt}}
\multiput(289.00,513.92)(0.588,-0.498){65}{\rule{0.571pt}{0.120pt}}
\multiput(289.00,514.17)(38.816,-34.000){2}{\rule{0.285pt}{0.400pt}}
\multiput(329.00,479.92)(1.298,-0.492){19}{\rule{1.118pt}{0.118pt}}
\multiput(329.00,480.17)(25.679,-11.000){2}{\rule{0.559pt}{0.400pt}}
\multiput(357.00,468.93)(1.484,-0.488){13}{\rule{1.250pt}{0.117pt}}
\multiput(357.00,469.17)(20.406,-8.000){2}{\rule{0.625pt}{0.400pt}}
\multiput(380.00,460.93)(1.935,-0.477){7}{\rule{1.540pt}{0.115pt}}
\multiput(380.00,461.17)(14.804,-5.000){2}{\rule{0.770pt}{0.400pt}}
\put(398,455.17){\rule{3.100pt}{0.400pt}}
\multiput(398.00,456.17)(8.566,-2.000){2}{\rule{1.550pt}{0.400pt}}
\multiput(413.00,453.95)(2.695,-0.447){3}{\rule{1.833pt}{0.108pt}}
\multiput(413.00,454.17)(9.195,-3.000){2}{\rule{0.917pt}{0.400pt}}
\put(426,450.67){\rule{2.891pt}{0.400pt}}
\multiput(426.00,451.17)(6.000,-1.000){2}{\rule{1.445pt}{0.400pt}}
\multiput(438.00,449.94)(1.358,-0.468){5}{\rule{1.100pt}{0.113pt}}
\multiput(438.00,450.17)(7.717,-4.000){2}{\rule{0.550pt}{0.400pt}}
\put(448,445.17){\rule{2.100pt}{0.400pt}}
\multiput(448.00,446.17)(5.641,-2.000){2}{\rule{1.050pt}{0.400pt}}
\multiput(458.00,443.93)(0.821,-0.477){7}{\rule{0.740pt}{0.115pt}}
\multiput(458.00,444.17)(6.464,-5.000){2}{\rule{0.370pt}{0.400pt}}
\put(466,438.17){\rule{1.700pt}{0.400pt}}
\multiput(466.00,439.17)(4.472,-2.000){2}{\rule{0.850pt}{0.400pt}}
\multiput(474.00,436.94)(1.066,-0.468){5}{\rule{0.900pt}{0.113pt}}
\multiput(474.00,437.17)(6.132,-4.000){2}{\rule{0.450pt}{0.400pt}}
\multiput(482.00,432.93)(0.599,-0.477){7}{\rule{0.580pt}{0.115pt}}
\multiput(482.00,433.17)(4.796,-5.000){2}{\rule{0.290pt}{0.400pt}}
\multiput(488.00,427.94)(0.920,-0.468){5}{\rule{0.800pt}{0.113pt}}
\multiput(488.00,428.17)(5.340,-4.000){2}{\rule{0.400pt}{0.400pt}}
\multiput(495.00,423.95)(2.248,-0.447){3}{\rule{1.567pt}{0.108pt}}
\multiput(495.00,424.17)(7.748,-3.000){2}{\rule{0.783pt}{0.400pt}}
\put(506,420.67){\rule{1.445pt}{0.400pt}}
\multiput(506.00,421.17)(3.000,-1.000){2}{\rule{0.723pt}{0.400pt}}
\multiput(512.59,418.26)(0.477,-0.710){7}{\rule{0.115pt}{0.660pt}}
\multiput(511.17,419.63)(5.000,-5.630){2}{\rule{0.400pt}{0.330pt}}
\put(517,412.17){\rule{1.100pt}{0.400pt}}
\multiput(517.00,413.17)(2.717,-2.000){2}{\rule{0.550pt}{0.400pt}}
\multiput(522.60,409.51)(0.468,-0.627){5}{\rule{0.113pt}{0.600pt}}
\multiput(521.17,410.75)(4.000,-3.755){2}{\rule{0.400pt}{0.300pt}}
\multiput(526.00,405.94)(1.212,-0.468){5}{\rule{1.000pt}{0.113pt}}
\multiput(526.00,406.17)(6.924,-4.000){2}{\rule{0.500pt}{0.400pt}}
\put(535,401.17){\rule{1.700pt}{0.400pt}}
\multiput(535.00,402.17)(4.472,-2.000){2}{\rule{0.850pt}{0.400pt}}
\put(543,399.67){\rule{0.964pt}{0.400pt}}
\multiput(543.00,400.17)(2.000,-1.000){2}{\rule{0.482pt}{0.400pt}}
\multiput(547.00,398.95)(0.462,-0.447){3}{\rule{0.500pt}{0.108pt}}
\multiput(547.00,399.17)(1.962,-3.000){2}{\rule{0.250pt}{0.400pt}}
\multiput(550.60,394.51)(0.468,-0.627){5}{\rule{0.113pt}{0.600pt}}
\multiput(549.17,395.75)(4.000,-3.755){2}{\rule{0.400pt}{0.300pt}}
\multiput(554.00,390.95)(0.462,-0.447){3}{\rule{0.500pt}{0.108pt}}
\multiput(554.00,391.17)(1.962,-3.000){2}{\rule{0.250pt}{0.400pt}}
\multiput(557.00,387.95)(0.462,-0.447){3}{\rule{0.500pt}{0.108pt}}
\multiput(557.00,388.17)(1.962,-3.000){2}{\rule{0.250pt}{0.400pt}}
\multiput(560.00,384.94)(0.920,-0.468){5}{\rule{0.800pt}{0.113pt}}
\multiput(560.00,385.17)(5.340,-4.000){2}{\rule{0.400pt}{0.400pt}}
\put(567,380.17){\rule{0.700pt}{0.400pt}}
\multiput(567.00,381.17)(1.547,-2.000){2}{\rule{0.350pt}{0.400pt}}
\put(570.17,374){\rule{0.400pt}{1.300pt}}
\multiput(569.17,377.30)(2.000,-3.302){2}{\rule{0.400pt}{0.650pt}}
\multiput(572.00,372.94)(0.774,-0.468){5}{\rule{0.700pt}{0.113pt}}
\multiput(572.00,373.17)(4.547,-4.000){2}{\rule{0.350pt}{0.400pt}}
\put(578,368.17){\rule{0.700pt}{0.400pt}}
\multiput(578.00,369.17)(1.547,-2.000){2}{\rule{0.350pt}{0.400pt}}
\put(581,366.17){\rule{1.100pt}{0.400pt}}
\multiput(581.00,367.17)(2.717,-2.000){2}{\rule{0.550pt}{0.400pt}}
\multiput(586.00,364.95)(0.685,-0.447){3}{\rule{0.633pt}{0.108pt}}
\multiput(586.00,365.17)(2.685,-3.000){2}{\rule{0.317pt}{0.400pt}}
\put(590,361.17){\rule{0.700pt}{0.400pt}}
\multiput(590.00,362.17)(1.547,-2.000){2}{\rule{0.350pt}{0.400pt}}
\multiput(593.00,359.93)(0.599,-0.477){7}{\rule{0.580pt}{0.115pt}}
\multiput(593.00,360.17)(4.796,-5.000){2}{\rule{0.290pt}{0.400pt}}
\multiput(599.61,352.82)(0.447,-0.909){3}{\rule{0.108pt}{0.767pt}}
\multiput(598.17,354.41)(3.000,-3.409){2}{\rule{0.400pt}{0.383pt}}
\multiput(602.60,348.09)(0.468,-0.774){5}{\rule{0.113pt}{0.700pt}}
\multiput(601.17,349.55)(4.000,-4.547){2}{\rule{0.400pt}{0.350pt}}
\multiput(606.00,343.95)(3.811,-0.447){3}{\rule{2.500pt}{0.108pt}}
\multiput(606.00,344.17)(12.811,-3.000){2}{\rule{1.250pt}{0.400pt}}
\multiput(624.00,340.94)(0.920,-0.468){5}{\rule{0.800pt}{0.113pt}}
\multiput(624.00,341.17)(5.340,-4.000){2}{\rule{0.400pt}{0.400pt}}
\multiput(631.00,336.95)(1.802,-0.447){3}{\rule{1.300pt}{0.108pt}}
\multiput(631.00,337.17)(6.302,-3.000){2}{\rule{0.650pt}{0.400pt}}
\put(639.67,332){\rule{0.400pt}{0.723pt}}
\multiput(639.17,333.50)(1.000,-1.500){2}{\rule{0.400pt}{0.361pt}}
\put(641.17,320){\rule{0.400pt}{2.500pt}}
\multiput(640.17,326.81)(2.000,-6.811){2}{\rule{0.400pt}{1.250pt}}
\multiput(643.60,315.85)(0.468,-1.212){5}{\rule{0.113pt}{1.000pt}}
\multiput(642.17,317.92)(4.000,-6.924){2}{\rule{0.400pt}{0.500pt}}
\multiput(647.00,309.92)(0.684,-0.492){19}{\rule{0.645pt}{0.118pt}}
\multiput(647.00,310.17)(13.660,-11.000){2}{\rule{0.323pt}{0.400pt}}
\multiput(662.60,297.09)(0.468,-0.774){5}{\rule{0.113pt}{0.700pt}}
\multiput(661.17,298.55)(4.000,-4.547){2}{\rule{0.400pt}{0.350pt}}
\put(666.17,288){\rule{0.400pt}{1.300pt}}
\multiput(665.17,291.30)(2.000,-3.302){2}{\rule{0.400pt}{0.650pt}}
\put(667.67,281){\rule{0.400pt}{1.686pt}}
\multiput(667.17,284.50)(1.000,-3.500){2}{\rule{0.400pt}{0.843pt}}
\multiput(669.59,278.37)(0.482,-0.671){9}{\rule{0.116pt}{0.633pt}}
\multiput(668.17,279.69)(6.000,-6.685){2}{\rule{0.400pt}{0.317pt}}
\multiput(675.59,270.37)(0.482,-0.671){9}{\rule{0.116pt}{0.633pt}}
\multiput(674.17,271.69)(6.000,-6.685){2}{\rule{0.400pt}{0.317pt}}
\multiput(681.59,262.51)(0.488,-0.626){13}{\rule{0.117pt}{0.600pt}}
\multiput(680.17,263.75)(8.000,-8.755){2}{\rule{0.400pt}{0.300pt}}
\put(689.17,244){\rule{0.400pt}{2.300pt}}
\multiput(688.17,250.23)(2.000,-6.226){2}{\rule{0.400pt}{1.150pt}}
\put(691.17,230){\rule{0.400pt}{2.900pt}}
\multiput(690.17,237.98)(2.000,-7.981){2}{\rule{0.400pt}{1.450pt}}
\multiput(693.58,227.37)(0.492,-0.669){21}{\rule{0.119pt}{0.633pt}}
\multiput(692.17,228.69)(12.000,-14.685){2}{\rule{0.400pt}{0.317pt}}
\multiput(705.58,210.10)(0.491,-1.069){17}{\rule{0.118pt}{0.940pt}}
\multiput(704.17,212.05)(10.000,-19.049){2}{\rule{0.400pt}{0.470pt}}
\multiput(715.00,191.92)(0.691,-0.497){55}{\rule{0.652pt}{0.120pt}}
\multiput(715.00,192.17)(38.647,-29.000){2}{\rule{0.326pt}{0.400pt}}
\put(755.17,113){\rule{0.400pt}{10.300pt}}
\multiput(754.17,142.62)(2.000,-29.622){2}{\rule{0.400pt}{5.150pt}}
\end{picture}

\noindent\small
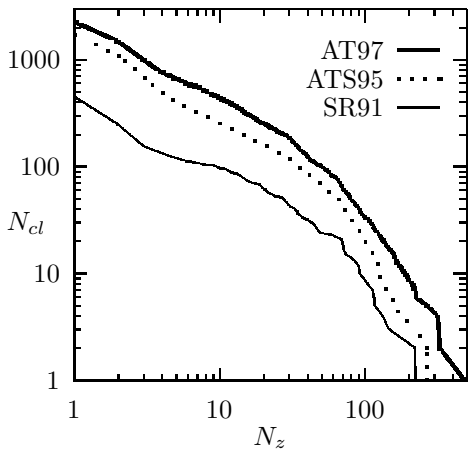
Figure 1:~The cumulative number \\
of clusters as a function of $ N_{z}$.
\end{minipage} \hfill
\begin{minipage}[b]{.46\linewidth}


\setlength{\unitlength}{0.240900pt}
\ifx\plotpoint\undefined\newsavebox{\plotpoint}\fi
\sbox{\plotpoint}{\rule[-0.200pt]{0.400pt}{0.400pt}}%
\begin{picture}(900,720)(180,0)
\font\gnuplot=cmr10 at 10pt
\gnuplot
\sbox{\plotpoint}{\rule[-0.200pt]{0.400pt}{0.400pt}}%
\put(220.0,113.0){\rule[-0.200pt]{148.394pt}{0.400pt}}
\put(220.0,113.0){\rule[-0.200pt]{0.400pt}{140.686pt}}
\put(220.0,113.0){\rule[-0.200pt]{4.818pt}{0.400pt}}
\put(198,113){\makebox(0,0)[r]{0}}
\put(816.0,113.0){\rule[-0.200pt]{4.818pt}{0.400pt}}
\put(220.0,191.0){\rule[-0.200pt]{4.818pt}{0.400pt}}
\put(198,191){\makebox(0,0)[r]{20}}
\put(816.0,191.0){\rule[-0.200pt]{4.818pt}{0.400pt}}
\put(220.0,269.0){\rule[-0.200pt]{4.818pt}{0.400pt}}
\put(198,269){\makebox(0,0)[r]{40}}
\put(816.0,269.0){\rule[-0.200pt]{4.818pt}{0.400pt}}
\put(220.0,347.0){\rule[-0.200pt]{4.818pt}{0.400pt}}
\put(816.0,347.0){\rule[-0.200pt]{4.818pt}{0.400pt}}
\put(220.0,424.0){\rule[-0.200pt]{4.818pt}{0.400pt}}
\put(816.0,424.0){\rule[-0.200pt]{4.818pt}{0.400pt}}
\put(220.0,502.0){\rule[-0.200pt]{4.818pt}{0.400pt}}
\put(198,502){\makebox(0,0)[r]{100}}
\put(816.0,502.0){\rule[-0.200pt]{4.818pt}{0.400pt}}
\put(220.0,580.0){\rule[-0.200pt]{4.818pt}{0.400pt}}
\put(198,580){\makebox(0,0)[r]{120}}
\put(816.0,580.0){\rule[-0.200pt]{4.818pt}{0.400pt}}
\put(220.0,658.0){\rule[-0.200pt]{4.818pt}{0.400pt}}
\put(198,658){\makebox(0,0)[r]{140}}
\put(816.0,658.0){\rule[-0.200pt]{4.818pt}{0.400pt}}
\put(220.0,113.0){\rule[-0.200pt]{0.400pt}{4.818pt}}
\put(220,68){\makebox(0,0){0}}
\put(220.0,677.0){\rule[-0.200pt]{0.400pt}{4.818pt}}
\put(348.0,113.0){\rule[-0.200pt]{0.400pt}{4.818pt}}
\put(348,68){\makebox(0,0){500}}
\put(348.0,677.0){\rule[-0.200pt]{0.400pt}{4.818pt}}
\put(477.0,113.0){\rule[-0.200pt]{0.400pt}{4.818pt}}
\put(477,68){\makebox(0,0){1000}}
\put(477.0,677.0){\rule[-0.200pt]{0.400pt}{4.818pt}}
\put(605.0,113.0){\rule[-0.200pt]{0.400pt}{4.818pt}}
\put(605,68){\makebox(0,0){1500}}
\put(605.0,677.0){\rule[-0.200pt]{0.400pt}{4.818pt}}
\put(733.0,113.0){\rule[-0.200pt]{0.400pt}{4.818pt}}
\put(733,68){\makebox(0,0){2000}}
\put(733.0,677.0){\rule[-0.200pt]{0.400pt}{4.818pt}}
\put(220.0,113.0){\rule[-0.200pt]{148.394pt}{0.400pt}}
\put(836.0,113.0){\rule[-0.200pt]{0.400pt}{140.686pt}}
\put(220.0,697.0){\rule[-0.200pt]{148.394pt}{0.400pt}}
\put(155,380){\makebox(0,0){\shortstack{$N_{\sigma}$}}}
\put(528,23){\makebox(0,0){\shortstack{$\sigma_V$~~[km/s]}}}
\put(220.0,113.0){\rule[-0.200pt]{0.400pt}{140.686pt}}
\sbox{\plotpoint}{\rule[-0.600pt]{1.200pt}{1.200pt}}%
\put(706,632){\makebox(0,0)[r]{All}}
\put(728.0,632.0){\rule[-0.600pt]{15.899pt}{1.200pt}}
\put(220,187){\usebox{\plotpoint}}
\put(220.0,187.0){\rule[-0.600pt]{12.286pt}{1.200pt}}
\put(271.0,187.0){\rule[-0.600pt]{1.200pt}{54.443pt}}
\put(271.0,413.0){\rule[-0.600pt]{12.527pt}{1.200pt}}
\put(323.0,413.0){\rule[-0.600pt]{1.200pt}{60.948pt}}
\put(323.0,666.0){\rule[-0.600pt]{12.286pt}{1.200pt}}
\put(374.0,666.0){\rule[-0.600pt]{1.200pt}{3.613pt}}
\put(374.0,681.0){\rule[-0.600pt]{12.286pt}{1.200pt}}
\put(425.0,623.0){\rule[-0.600pt]{1.200pt}{13.972pt}}
\put(425.0,623.0){\rule[-0.600pt]{12.527pt}{1.200pt}}
\put(477.0,315.0){\rule[-0.600pt]{1.200pt}{74.197pt}}
\put(477.0,315.0){\rule[-0.600pt]{12.286pt}{1.200pt}}
\put(528.0,203.0){\rule[-0.600pt]{1.200pt}{26.981pt}}
\put(528.0,203.0){\rule[-0.600pt]{12.286pt}{1.200pt}}
\put(579.0,156.0){\rule[-0.600pt]{1.200pt}{11.322pt}}
\put(579.0,156.0){\rule[-0.600pt]{12.527pt}{1.200pt}}
\put(631.0,132.0){\rule[-0.600pt]{1.200pt}{5.782pt}}
\put(631.0,132.0){\rule[-0.600pt]{12.286pt}{1.200pt}}
\put(682.0,132.0){\usebox{\plotpoint}}
\put(682.0,136.0){\rule[-0.600pt]{12.286pt}{1.200pt}}
\put(733.0,113.0){\rule[-0.600pt]{1.200pt}{5.541pt}}
\put(733.0,113.0){\rule[-0.600pt]{24.813pt}{1.200pt}}
\put(836.0,113.0){\usebox{\plotpoint}}
\sbox{\plotpoint}{\rule[-0.200pt]{0.400pt}{0.400pt}}%
\put(706,587){\makebox(0,0)[r]{$N >= 10$}}
\put(728.0,587.0){\rule[-0.200pt]{15.899pt}{0.400pt}}
\put(220,132){\usebox{\plotpoint}}
\put(220.0,132.0){\rule[-0.200pt]{12.286pt}{0.400pt}}
\put(271.0,132.0){\rule[-0.200pt]{0.400pt}{42.398pt}}
\put(271.0,308.0){\rule[-0.200pt]{12.527pt}{0.400pt}}
\put(323.0,308.0){\rule[-0.200pt]{0.400pt}{44.085pt}}
\put(323.0,491.0){\rule[-0.200pt]{12.286pt}{0.400pt}}
\put(374.0,491.0){\rule[-0.200pt]{0.400pt}{8.431pt}}
\put(374.0,526.0){\rule[-0.200pt]{12.286pt}{0.400pt}}
\put(425.0,498.0){\rule[-0.200pt]{0.400pt}{6.745pt}}
\put(425.0,498.0){\rule[-0.200pt]{12.527pt}{0.400pt}}
\put(477.0,265.0){\rule[-0.200pt]{0.400pt}{56.130pt}}
\put(477.0,265.0){\rule[-0.200pt]{12.286pt}{0.400pt}}
\put(528.0,171.0){\rule[-0.200pt]{0.400pt}{22.645pt}}
\put(528.0,171.0){\rule[-0.200pt]{12.286pt}{0.400pt}}
\put(579.0,140.0){\rule[-0.200pt]{0.400pt}{7.468pt}}
\put(579.0,140.0){\rule[-0.200pt]{12.527pt}{0.400pt}}
\put(631.0,132.0){\rule[-0.200pt]{0.400pt}{1.927pt}}
\put(631.0,132.0){\rule[-0.200pt]{24.572pt}{0.400pt}}
\put(733.0,113.0){\rule[-0.200pt]{0.400pt}{4.577pt}}
\put(733.0,113.0){\rule[-0.200pt]{24.813pt}{0.400pt}}
\sbox{\plotpoint}{\rule[-0.400pt]{0.800pt}{0.800pt}}%
\put(706,542){\makebox(0,0)[r]{S-clusters}}
\put(728.0,542.0){\rule[-0.400pt]{15.899pt}{0.800pt}}
\put(220,148){\usebox{\plotpoint}}
\put(220.0,148.0){\rule[-0.400pt]{12.286pt}{0.800pt}}
\put(271.0,148.0){\rule[-0.400pt]{0.800pt}{19.754pt}}
\put(271.0,230.0){\rule[-0.400pt]{12.527pt}{0.800pt}}
\put(323.0,230.0){\rule[-0.400pt]{0.800pt}{0.964pt}}
\put(323.0,234.0){\rule[-0.400pt]{12.286pt}{0.800pt}}
\put(374.0,234.0){\rule[-0.400pt]{0.800pt}{5.541pt}}
\put(374.0,257.0){\rule[-0.400pt]{12.286pt}{0.800pt}}
\put(425.0,226.0){\rule[-0.400pt]{0.800pt}{7.468pt}}
\put(425.0,226.0){\rule[-0.400pt]{12.527pt}{0.800pt}}
\put(477.0,168.0){\rule[-0.400pt]{0.800pt}{13.972pt}}
\put(477.0,168.0){\rule[-0.400pt]{12.286pt}{0.800pt}}
\put(528.0,140.0){\rule[-0.400pt]{0.800pt}{6.745pt}}
\put(528.0,140.0){\rule[-0.400pt]{12.286pt}{0.800pt}}
\put(579.0,125.0){\rule[-0.400pt]{0.800pt}{3.613pt}}
\put(579.0,125.0){\rule[-0.400pt]{12.527pt}{0.800pt}}
\put(631.0,121.0){\rule[-0.400pt]{0.800pt}{0.964pt}}
\put(631.0,121.0){\rule[-0.400pt]{12.286pt}{0.800pt}}
\put(682.0,113.0){\rule[-0.400pt]{0.800pt}{1.927pt}}
\put(682.0,113.0){\rule[-0.400pt]{37.099pt}{0.800pt}}
\end{picture}

\noindent\small
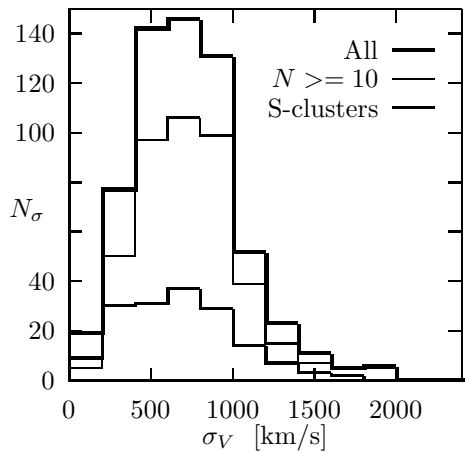
\hspace*{4mm} Figure 2:~The distribution of cluster \\
\hspace*{4mm} velocity dispersions for 3 samples.
\end{minipage}
\normalsize

\vspace*{2mm}
With the increasing number of
redshift surveys there are more and more cases where galaxy data from
different sources for the same cluster need to be merged.  In numerous
cases this was impossible since no individual galaxy redshift data were
published. As a result there are 15 clusters in our compilation with
N$_{z}\ge$10 and no $\sigma_V$ available. For clusters with large N$_{z}$ 
and more than two papers to merge, we often chose the paper with 
the largest N$_{z}$ but quoted the additional references in the notes. 
It would be very helpful if authors of cluster redshifts included
both, positions and velocities of individual galaxies, and clearly state
whether velocities are geo-, helio- or galactocentric.
Ideally, these data should always and immediately be integrated in
a complete galaxy redshift compilation. If possible we incorporate 
data from most recent preprints (e.g.~from the LANL/SISSA servers).

For the rich ACO clusters (A-names) we use the redshift estimate
in~\cite{pw92}. For the supplementary clusters
we use the function proposed for the rich southern clusters by
ACO~\cite{aco89}, but scaled down by 30\% (as found by one of
us~\cite{and91}). We confirm the latter scaling with twice the number
of S-clusters with measured z available now.
We also find that the S-clusters do suffer from
line-of-sight superpositions more than the A-clusters:
7.6\% of the distinct S-cluster names appear with more than one
entry, versus 5.6\% for the distinct A-clusters.

In Fig.\,3 we plot the distribution of estimated redshifts of
all ACO- and S-clusters on top of that of the measured redshifts.
Measured redshifts (N$_{z}>$0) are now available
for all northern clusters with z$_{\rm est}<$0.076, for all southern
A-clusters with z$_{\rm est}<$0.057 and for all S-clusters with
z$_{\rm est}<$0.035.

\setlength{\unitlength}{0.240900pt}
\ifx\plotpoint\undefined\newsavebox{\plotpoint}\fi
\sbox{\plotpoint}{\rule[-0.200pt]{0.400pt}{0.400pt}}%
\begin{picture}(1349,675)(80,0)
\font\gnuplot=cmr10 at 10pt
\gnuplot
\sbox{\plotpoint}{\rule[-0.200pt]{0.400pt}{0.400pt}}%
\put(220.0,113.0){\rule[-0.200pt]{256.558pt}{0.400pt}}
\put(220.0,113.0){\rule[-0.200pt]{0.400pt}{129.845pt}}
\put(220.0,113.0){\rule[-0.200pt]{4.818pt}{0.400pt}}
\put(198,113){\makebox(0,0)[r]{0}}
\put(1265.0,113.0){\rule[-0.200pt]{4.818pt}{0.400pt}}
\put(220.0,173.0){\rule[-0.200pt]{4.818pt}{0.400pt}}
\put(1265.0,173.0){\rule[-0.200pt]{4.818pt}{0.400pt}}
\put(220.0,233.0){\rule[-0.200pt]{4.818pt}{0.400pt}}
\put(198,233){\makebox(0,0)[r]{200}}
\put(1265.0,233.0){\rule[-0.200pt]{4.818pt}{0.400pt}}
\put(220.0,293.0){\rule[-0.200pt]{4.818pt}{0.400pt}}
\put(1265.0,293.0){\rule[-0.200pt]{4.818pt}{0.400pt}}
\put(220.0,353.0){\rule[-0.200pt]{4.818pt}{0.400pt}}
\put(198,353){\makebox(0,0)[r]{400}}
\put(1265.0,353.0){\rule[-0.200pt]{4.818pt}{0.400pt}}
\put(220.0,412.0){\rule[-0.200pt]{4.818pt}{0.400pt}}
\put(1265.0,412.0){\rule[-0.200pt]{4.818pt}{0.400pt}}
\put(220.0,472.0){\rule[-0.200pt]{4.818pt}{0.400pt}}
\put(198,472){\makebox(0,0)[r]{600}}
\put(1265.0,472.0){\rule[-0.200pt]{4.818pt}{0.400pt}}
\put(220.0,532.0){\rule[-0.200pt]{4.818pt}{0.400pt}}
\put(1265.0,532.0){\rule[-0.200pt]{4.818pt}{0.400pt}}
\put(220.0,592.0){\rule[-0.200pt]{4.818pt}{0.400pt}}
\put(198,592){\makebox(0,0)[r]{800}}
\put(1265.0,592.0){\rule[-0.200pt]{4.818pt}{0.400pt}}
\put(220.0,652.0){\rule[-0.200pt]{4.818pt}{0.400pt}}
\put(1265.0,652.0){\rule[-0.200pt]{4.818pt}{0.400pt}}
\put(220.0,113.0){\rule[-0.200pt]{0.400pt}{4.818pt}}
\put(220,68){\makebox(0,0){0}}
\put(220.0,632.0){\rule[-0.200pt]{0.400pt}{4.818pt}}
\put(353.0,113.0){\rule[-0.200pt]{0.400pt}{4.818pt}}
\put(353,68){\makebox(0,0){0.05}}
\put(353.0,632.0){\rule[-0.200pt]{0.400pt}{4.818pt}}
\put(486.0,113.0){\rule[-0.200pt]{0.400pt}{4.818pt}}
\put(486,68){\makebox(0,0){0.1}}
\put(486.0,632.0){\rule[-0.200pt]{0.400pt}{4.818pt}}
\put(619.0,113.0){\rule[-0.200pt]{0.400pt}{4.818pt}}
\put(619,68){\makebox(0,0){0.15}}
\put(619.0,632.0){\rule[-0.200pt]{0.400pt}{4.818pt}}
\put(753.0,113.0){\rule[-0.200pt]{0.400pt}{4.818pt}}
\put(753,68){\makebox(0,0){0.2}}
\put(753.0,632.0){\rule[-0.200pt]{0.400pt}{4.818pt}}
\put(886.0,113.0){\rule[-0.200pt]{0.400pt}{4.818pt}}
\put(886,68){\makebox(0,0){0.25}}
\put(886.0,632.0){\rule[-0.200pt]{0.400pt}{4.818pt}}
\put(1019.0,113.0){\rule[-0.200pt]{0.400pt}{4.818pt}}
\put(1019,68){\makebox(0,0){0.3}}
\put(1019.0,632.0){\rule[-0.200pt]{0.400pt}{4.818pt}}
\put(1152.0,113.0){\rule[-0.200pt]{0.400pt}{4.818pt}}
\put(1152,68){\makebox(0,0){0.35}}
\put(1152.0,632.0){\rule[-0.200pt]{0.400pt}{4.818pt}}
\put(1285.0,113.0){\rule[-0.200pt]{0.400pt}{4.818pt}}
\put(1285,68){\makebox(0,0){0.4}}
\put(1285.0,632.0){\rule[-0.200pt]{0.400pt}{4.818pt}}
\put(220.0,113.0){\rule[-0.200pt]{256.558pt}{0.400pt}}
\put(1285.0,113.0){\rule[-0.200pt]{0.400pt}{129.845pt}}
\put(220.0,652.0){\rule[-0.200pt]{256.558pt}{0.400pt}}
\put(95,405){\makebox(0,0){$N_z$}}
\put(752,23){\makebox(0,0){z}}
\put(220.0,113.0){\rule[-0.200pt]{0.400pt}{129.845pt}}
\put(1155,587){\makebox(0,0)[r]{all (estimated)}}
\put(1177.0,587.0){\rule[-0.200pt]{15.899pt}{0.400pt}}
\put(220,113){\usebox{\plotpoint}}
\put(220.0,113.0){\rule[-0.200pt]{0.400pt}{3.854pt}}
\put(220.0,129.0){\rule[-0.200pt]{12.768pt}{0.400pt}}
\put(273.0,129.0){\rule[-0.200pt]{0.400pt}{18.067pt}}
\put(273.0,204.0){\rule[-0.200pt]{13.009pt}{0.400pt}}
\put(327.0,204.0){\rule[-0.200pt]{0.400pt}{38.785pt}}
\put(327.0,365.0){\rule[-0.200pt]{12.768pt}{0.400pt}}
\put(380.0,365.0){\rule[-0.200pt]{0.400pt}{19.031pt}}
\put(380.0,444.0){\rule[-0.200pt]{12.768pt}{0.400pt}}
\put(433.0,362.0){\rule[-0.200pt]{0.400pt}{19.754pt}}
\put(433.0,362.0){\rule[-0.200pt]{12.768pt}{0.400pt}}
\put(486.0,362.0){\rule[-0.200pt]{0.400pt}{23.608pt}}
\put(486.0,460.0){\rule[-0.200pt]{12.768pt}{0.400pt}}
\put(539.0,460.0){\rule[-0.200pt]{0.400pt}{36.376pt}}
\put(539.0,611.0){\rule[-0.200pt]{13.009pt}{0.400pt}}
\put(593.0,550.0){\rule[-0.200pt]{0.400pt}{14.695pt}}
\put(593.0,550.0){\rule[-0.200pt]{12.768pt}{0.400pt}}
\put(646.0,535.0){\rule[-0.200pt]{0.400pt}{3.613pt}}
\put(646.0,535.0){\rule[-0.200pt]{12.768pt}{0.400pt}}
\put(699.0,414.0){\rule[-0.200pt]{0.400pt}{29.149pt}}
\put(699.0,414.0){\rule[-0.200pt]{13.009pt}{0.400pt}}
\put(753.0,265.0){\rule[-0.200pt]{0.400pt}{35.894pt}}
\put(753.0,265.0){\rule[-0.200pt]{12.768pt}{0.400pt}}
\put(806.0,178.0){\rule[-0.200pt]{0.400pt}{20.958pt}}
\put(806.0,178.0){\rule[-0.200pt]{12.768pt}{0.400pt}}
\put(859.0,154.0){\rule[-0.200pt]{0.400pt}{5.782pt}}
\put(859.0,154.0){\rule[-0.200pt]{12.768pt}{0.400pt}}
\put(912.0,124.0){\rule[-0.200pt]{0.400pt}{7.227pt}}
\put(912.0,124.0){\rule[-0.200pt]{13.009pt}{0.400pt}}
\put(966.0,120.0){\rule[-0.200pt]{0.400pt}{0.964pt}}
\put(966.0,120.0){\rule[-0.200pt]{12.768pt}{0.400pt}}
\put(1019.0,118.0){\rule[-0.200pt]{0.400pt}{0.482pt}}
\put(1019.0,118.0){\rule[-0.200pt]{12.768pt}{0.400pt}}
\put(1072.0,114.0){\rule[-0.200pt]{0.400pt}{0.964pt}}
\put(1072.0,114.0){\rule[-0.200pt]{25.535pt}{0.400pt}}
\put(1178.0,113.0){\usebox{\plotpoint}}
\put(1178.0,113.0){\rule[-0.200pt]{25.776pt}{0.400pt}}
\sbox{\plotpoint}{\rule[-0.600pt]{1.200pt}{1.200pt}}%
\put(1155,542){\makebox(0,0)[r]{observed}}
\put(1177.0,542.0){\rule[-0.600pt]{15.899pt}{1.200pt}}
\put(220,113){\usebox{\plotpoint}}
\put(220.0,113.0){\rule[-0.600pt]{1.200pt}{4.577pt}}
\put(220.0,132.0){\rule[-0.600pt]{12.768pt}{1.200pt}}
\put(273.0,132.0){\rule[-0.600pt]{1.200pt}{16.863pt}}
\put(273.0,202.0){\rule[-0.600pt]{13.009pt}{1.200pt}}
\put(327.0,202.0){\rule[-0.600pt]{1.200pt}{22.645pt}}
\put(327.0,296.0){\rule[-0.600pt]{12.768pt}{1.200pt}}
\put(380.0,296.0){\rule[-0.600pt]{1.200pt}{15.418pt}}
\put(380.0,360.0){\rule[-0.600pt]{12.768pt}{1.200pt}}
\put(433.0,293.0){\rule[-0.600pt]{1.200pt}{16.140pt}}
\put(433.0,293.0){\rule[-0.600pt]{12.768pt}{1.200pt}}
\put(486.0,266.0){\rule[-0.600pt]{1.200pt}{6.504pt}}
\put(486.0,266.0){\rule[-0.600pt]{12.768pt}{1.200pt}}
\put(539.0,238.0){\rule[-0.600pt]{1.200pt}{6.745pt}}
\put(539.0,238.0){\rule[-0.600pt]{13.009pt}{1.200pt}}
\put(593.0,211.0){\rule[-0.600pt]{1.200pt}{6.504pt}}
\put(593.0,211.0){\rule[-0.600pt]{12.768pt}{1.200pt}}
\put(646.0,183.0){\rule[-0.600pt]{1.200pt}{6.745pt}}
\put(646.0,183.0){\rule[-0.600pt]{12.768pt}{1.200pt}}
\put(699.0,175.0){\rule[-0.600pt]{1.200pt}{1.927pt}}
\put(699.0,175.0){\rule[-0.600pt]{13.009pt}{1.200pt}}
\put(753.0,158.0){\rule[-0.600pt]{1.200pt}{4.095pt}}
\put(753.0,158.0){\rule[-0.600pt]{12.768pt}{1.200pt}}
\put(806.0,144.0){\rule[-0.600pt]{1.200pt}{3.373pt}}
\put(806.0,144.0){\rule[-0.600pt]{12.768pt}{1.200pt}}
\put(859.0,130.0){\rule[-0.600pt]{1.200pt}{3.373pt}}
\put(859.0,130.0){\rule[-0.600pt]{12.768pt}{1.200pt}}
\put(912.0,117.0){\rule[-0.600pt]{1.200pt}{3.132pt}}
\put(912.0,117.0){\rule[-0.600pt]{13.009pt}{1.200pt}}
\put(966.0,117.0){\rule[-0.600pt]{1.200pt}{1.686pt}}
\put(966.0,124.0){\rule[-0.600pt]{12.768pt}{1.200pt}}
\put(1019.0,118.0){\rule[-0.600pt]{1.200pt}{1.445pt}}
\put(1019.0,118.0){\rule[-0.600pt]{12.768pt}{1.200pt}}
\put(1072.0,117.0){\usebox{\plotpoint}}
\put(1072.0,117.0){\rule[-0.600pt]{12.768pt}{1.200pt}}
\put(1125.0,115.0){\usebox{\plotpoint}}
\put(1125.0,115.0){\rule[-0.600pt]{25.776pt}{1.200pt}}
\put(1232.0,113.0){\usebox{\plotpoint}}
\put(1232.0,113.0){\rule[-0.600pt]{12.768pt}{1.200pt}}
\end{picture}

\vspace*{-2mm}
\begin{center}
\small
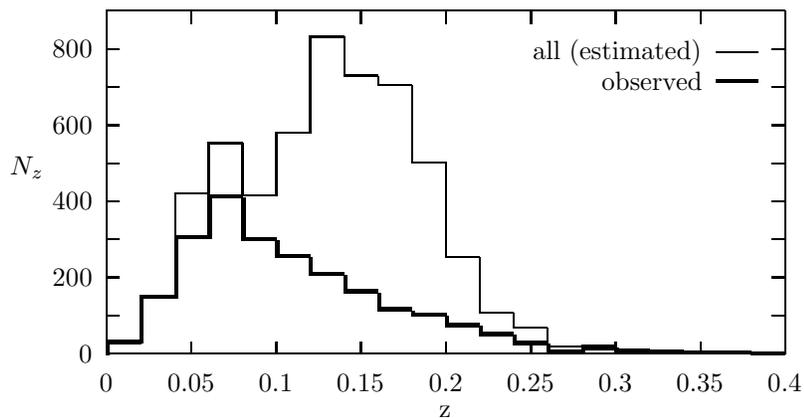
Fig.~3.~~Redshift distribution of measured clusters (z$_{obs}$) 
and all (A+S) clusters 
\end{center}

\section{Applications}

We used our compilation to establish
the currently most complete catalog of superclusters of galaxies~\cite{svn1}.
Studying the 3-dimensional distribution of these superclusters we found that
the richest superclusters occupy a more
or less regular lattice of $\sim$120\,Mpc spacing~\cite{natpap}. 
We see about five periods of this lattice. The observed regularity 
is in conflict with current models of structure formation.
Most recently the data have been used to derive the cluster correlation
function and the power spectrum~\cite{svn2,svn3}.
At present we are studying the shapes of
rich superclusters (Jaaniste et al., in prep.).

\medskip
\noindent
\ \ \ The authors are grateful to the organizers for covering all local expenses.

\end{document}